\documentclass{desyproc}

\begin{document}
\title{First sensitivity limits of the ALPS TES detector}

\author{{\slshape Jan Dreyling-Eschweiler$^1$, for the ALPS-II collaboration}\\[1ex]
$^1$Deutsches Elektronen-Synchrotron (DESY), Hamburg, Germany}

\contribID{dreyling-eschweiler\_jan}

\desyproc{DESY-PROC-2014-XX}
\acronym{Patras 2014} 
\doi  

\maketitle

\begin{abstract}
The Any Light Particle Search II (ALPS II) requires a sensitive detection of 1064~nm photons. 
Thus, a low dark count rate ($DC$) and a high detection efficiency ($DE$) is needed.
ALPS has set up a transition-edge sensor (TES) detector system, namely the ALPS TES detector. 
It is found that thermal photons from room temperature surfaces are the main contribution of dark counts for 1064~nm photon signals.
Furthermore, the current setup of the ALPS TES detector shows an improvement compared to using the ALPS I detector.
\end{abstract}

\section{Introduction} 

ALPS II is a light-shining-through-a-wall experiment based on an optical laser \cite{januschek2014}.  
Compared to ALPS I, the overall experimental sensitivity will be increased by extending the magnetic length, improving the optical system and setting up a more sensitive photon detector.%
\footnote{These improvements can be illustrated by the board game ``Axionator'', see the corresponding slides or contact the author.}
According to the latter, a detection of low-flux single 1064~nm photons is required \cite{tdr2013}.
Therefore, we have set to work on developing a TES detector system \cite{de2013}.  
In the last year, the completed setup has been characterized concerning its experimental sensitivity for the first time \cite{de2014phd}.

 \section{ALPS detector sensitivity}

The ALPS II experiment asks for a detection of low rates of single 1064~nm photons. The sensitivity of the axion-like particle-photon coupling, $g_{ {\rm a}\gamma}$, concerning only the ALPS detector is defined as \cite{seggern2013}:
\begin{eqnarray}
  \mathcal{S}_{\rm detector}(g_{ {\rm a}\gamma}) = \left({\sqrt{DC}}/{DE}\right)^{1/4} 
  \label{eqn:sens}
\end{eqnarray}
where $DC$ is the dark count%
\footnote{A dark count is defined as an event which cannot be distinguished from signals.}
rate and $DE$ is the (overall) detection efficiency of the detector.
The smaller this figure of merit the higher the gain for the ALPS experiment.

The ALPS I detector was a commercial state-of-the-art silicon-based charged-coupled device (CCD) \cite{alps2011}. 
Using this CCD camera to detect 1064~nm light, the sensitivity is limited due to $DE_{\rm CCD} = 1.2$~\%  \cite{seggern2014phd}.
With $DC_{\rm CCD} = 1.2 \cdot 10^{-3}$~s$^{-1}$ \cite{seggern2014phd} it results: 
\begin{eqnarray}
\mathcal{S}_{\rm CCD} = 1.303 \,\, {{\rm s}^{-1/8}}
  \label{eqn:sens:ccd}
\end{eqnarray}
The goal of a new detector is to reach a higher sensitivity or lower figure of merit, $\mathcal{S}_{\rm detector}$.

\section{The ALPS TES detector}

Detectors based on transition-edge sensors (TESs) can reach a high sensitivity \cite{irwin2005}. TESs are operated within the superconducting transition of the sensor's material. 
Thus, small temperature changes like by a photon cause measurable changes of the electrical resistance. 

For the ALPS TES detector, we are using a tungsten-based TES from NIST \cite{lita2010}. 
These kind of sensors are optimized for detecting single photons around 1064~nm.
Single mode fibers are attached to NIST TES chips \cite{miller2012}. 
For 1064~nm signals, a detection efficiency  of $DE = $ 97~\% $\pm$ 2~\% syst. $\pm$ 1~\% stat. was measured \cite{comm:lita}.
Exact measurements of ALPS TES detector's $DE$ are underway within the collaboration. 
For the considered dark count rate in this paper (Sec.~\ref{sec:dc}), we conservatively estimate  \cite{de2014phd}
\begin{eqnarray}
DE_{\rm TES} = 18 ~\% 
\label{eqn:de}
\end{eqnarray}
which includes a low analysis efficiency due to necessary cuts on the signal region (Sec.~\ref{sec:dc}).

We are operating the NIST TES at a bath temperature of $T_{\rm bath} = 80$~mK $\pm~25$~$\mu$K due to the superconducting transition $T_{\rm c} \approx 140$~mK of the TES \cite{de2014phd}. 
The bath is provided by an adiabatic demagnetization refrigerator (ADR) system. 
The 2.5~K-precooling is provided by a pulse tube cooler. 
The subsequent mK-cooling is provided by a 6~T magnet and a salt pills unit, which allows to establish a continuous 80 mK-operation for $>20$~h.
The low-noise current read-out of the TES circuit is provided by an dc 2-stage SQUID from PTB \cite{drung2007}.
For signal calibration as well as for long-background measurements, we have used an oscilloscope DPO7104C from Tektronix as data acquisition system (DAQ).

\section{Signals and thermal photonic background}
\label{sec:signal}

\begin{wrapfigure}{r}{0.6\textwidth}
  \centering
  \includegraphics[width=0.57\textwidth]{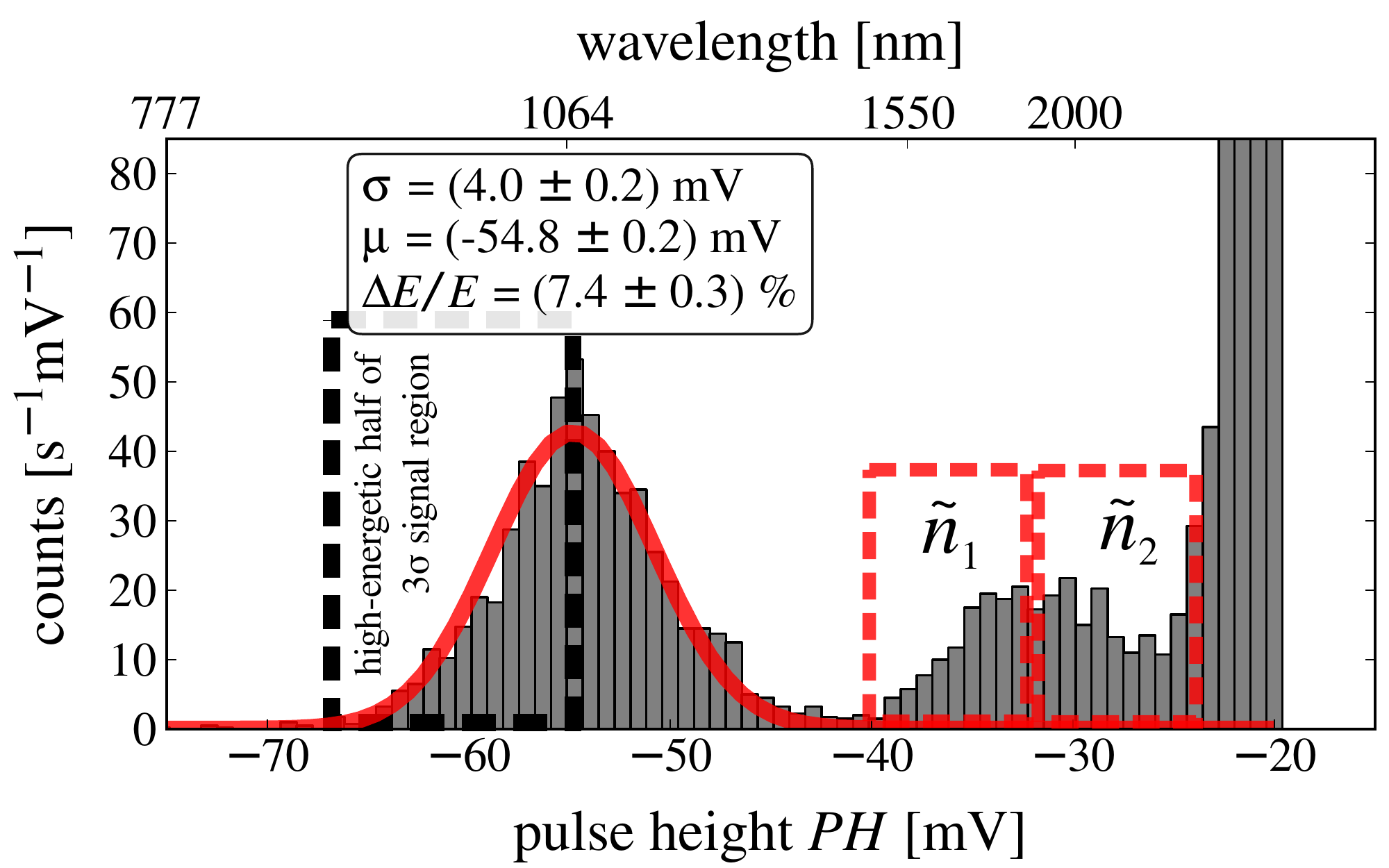}
  \caption{Exemplary pulse height distribution: Pulse heights $PH$ of trigger events are negatively plotted in mV-units of the DAQ (bottom x-axis) and in nm-units of wavelength (top x-axis). Signal photons are around 1064~nm (red Gaussian shape). For the results, the high-energetic half of the 3$\sigma$-region is considered (dashed black box). Sensor noise is distributed around 0~mV. 
  Thermal photons from room temperature surfaces are roughly found between 1550 and 2000~nm. For a pile-up estimate, the thermal peak is split in two contributions $\tilde{n}_1$ and $\tilde{n}_2$ (dashed red boxes).
}
  \label{fig:phd}
\end{wrapfigure}

Applying an attenuated laser for single photon source,
we consider the distribution of the pulse heights $PH$ (Fig.~\ref{fig:phd}).
We observe three event classes \cite{de2014phd}:
\begin{itemize}
  \item Signal  photons: 1064~nm photons normally distribute around $PH \approx -55$~mV.
    Fitting a Gaussian distribution allows to define a signal region and to determine an energy resolution of a 1064~nm photon signal. We observe $(\Delta E/E)_{\rm 1064~nm} = 7-8~\%$.
  \item Sensor noise: The signal peak is clearly separated from the noise peak around $PH = 0$~mV.
    The noise is mainly caused by the fundamental TES noise. 
  \item Thermal photons: Thermal photon events are observed due to the warm fiber end at room temperature. The optical transmittance of the single mode fiber and the TES absorptivity cause an effective peak between 1550 and 2000~nm. 
Thermal photons through the fiber are confirmed by operating the TES in a $<4$~K-environment without optical fiber link.
Thermal photons are also observed by other groups as a reasonable background source \cite{miller2007}.
\end{itemize}

While the intrinsic noise is no possible source for dark counts of 1064~nm signals, the spectrum of thermal photons can reach the 1064~nm signal region and result in a dark count. Further background events, like intrinsic, are well discriminated by a proper pulse shape analysis  \cite{de2014phd}.
To estimate an expected dark count rate, we consider a conservative model:
300~K black body spectrum, 
no optical losses and an 
overall energy resolution of $\Delta E/E = 10~\%$.
Furthermore, we only consider the high-energetic 3$\sigma$-region of 1064~nm signals which corresponds to photon wavelength between 818 and 1064~nm (Sec.~\ref{sec:dc}). 
This results in a photon rate of $3.4 \cdot 10^{-4}$~s$^{-1}$ due to a room-temperature black body spectrum. This number is compared to long-term measurements in the next section.

\section{Dark count rate}
\label{sec:dc}

To determine the dark count rate of a fiber-coupled TES, the warm fiber end outside the cryostat was adequately covered, so that no ambient light can couple into the open fiber end. 
We have analyzed each trigger event of two long-term measurements ($> 14$~h) offline. 
Only events showing a single photon pulse shape were taken into account \cite{de2014phd}.

Furthermore, as mentioned above, we only consider the high-energetic half 3$\sigma$-region of 1064~nm signals due to the used trigger level.
This is necessary because of the used DAQ: A corresponding low trigger level around a $PH$ of 1550~nm photons results in a too high trigger rate due to the thermal spectrum, which would cause dead time effects.
For this setup, we determine a dark count rate of
\begin{eqnarray}
  DC_{\rm TES} =  8.6 \cdot 10^{-3} \,\, {\rm s}^{-1}
\label{eqn:dc}
\end{eqnarray}

Comparing this rate to the estimate rate due to a black body spectrum (Sec. \ref{sec:signal}) the measured rate is $\sim$1.5 orders of magnitude higher than the expected rate. 
It is found that a plausible explanation are pile-up events of thermal photons \cite{de2014phd}. 
A first-order pile-up rate $\tilde{n}_{\rm eff}$  can be estimated by a formula describing accidental coincidences \cite{eckart1938}:
$\tilde{n}_{\rm eff} = 2\, \tau \, \tilde{n}_1 \, \tilde{n}_2$
where $\tau$ is the resolving time and $\tilde{n}_1$ and $\tilde{n}_2$ uncorrelated rates.
The time resolution of the current setup is constrained by the analysis method and is conservatively estimated by $\tau \approx 0.5$~$\mu$s. 
The rates $\tilde{n}_1$ and $\tilde{n}_2$ are estimated from Fig.~\ref{fig:phd}: We split the thermal peak in two parts and assume that two photons from each part can effectively combine to a 1064~nm signal event.
It is $\tilde{n}_1 \approx \tilde{n}_2 \approx 10^{2}~{\rm s}^{-1}$. Thus, the rate for first-order pile-ups imitating 1064 nm events is $\tilde{n}_{\rm eff} \simeq 10^{-2}~{\rm s}^{-1}$. 
$\tilde{n}_{\rm eff}$ is in the same order of magnitude as the measured rate $DC_{\rm TES}$ (Eq.~\ref{eqn:dc}). 

\section{Conclusion}

Using the ALPS TES detector, the experimental sensitivity (Eq.~\ref{eqn:sens}) results in
\begin{eqnarray}
\mathcal{S}_{\rm TES} = 0.847 \,\, {{\rm s}^{-1/8}}
  \label{eqn:sens:tes}
\end{eqnarray}
where we used the estimated $DE_{\rm TES}$ (Eq.~\ref{eqn:de}) and the measured $DC_{\rm TES}$ (Eq.~\ref{eqn:dc}).
Compared to the CCD (Eq.~\ref{eqn:sens:ccd}), this results in a sensitivity gain of $\mathcal{S}_{\rm CCD}/\mathcal{S}_{\rm TES} = 1.54$ for the ALPS experiment.
This proves the gain of using a TES-based detector instead of the CCD camera system to detect single 1064 nm photons.
Furthermore, we are motivated to improve the ALPS TES detector. Most promising is to reduce the thermal background by using an optical bandpass filter and an improved analysis for pile-up rejection.

\section*{Acknowledgments}

The ALPS collaboration wants to thank the PTB and NIST for the support of their superconducting devices.
 

\begin{footnotesize}


\end{footnotesize}


\end{document}